\documentclass{sigchi}
\pdfoutput=1
\usepackage{balance}       
\usepackage{graphics}      
\usepackage[T1]{fontenc}   
\usepackage{txfonts}
\usepackage{mathptmx}
\usepackage[pdflang={en-US},pdftex]{hyperref}
\usepackage{color}
\usepackage{booktabs}
\usepackage{textcomp}

\usepackage{microtype}        
\usepackage{ccicons}          

\usepackage{todonotes}

\def\plaintitle{Novelty Learning via Collaborative Proximity Filtering}

\def\emptyauthor{}
\def\plainkeywords{Recommender Systems; User Behaviors; Boredom; Novelty; User Preferences; Implicit Preferences; Latent Tastes}

\makeatletter
\def\url@leostyle{%
  \@ifundefined{selectfont}{
    \def\UrlFont{\sf}
  }{
    \def\UrlFont{\small\bf\ttfamily}
  }}
\def\@copyrightspace{\relax} 
\makeatother
\def\pprw{8.5in}
\def\pprh{11in}

\definecolor{linkColor}{RGB}{6,125,233}
\hypersetup{%
  pdftitle={\plaintitle},
  pdfauthor={\emptyauthor},
  pdfkeywords={\plainkeywords},
  pdfdisplaydoctitle=true, 
  bookmarksnumbered,
  pdfstartview={FitH},
  colorlinks,
  citecolor=black,
  filecolor=black,
  linkcolor=black,
  urlcolor=linkColor,
  breaklinks=true,
  hypertexnames=false
}

\usepackage{standalone}
\usepackage{tikz}
\usetikzlibrary{arrows,calc,shapes,decorations.pathreplacing,backgrounds,automata,positioning,shadows,fadings}
\tikzset{
    >=stealth',
    punkt/.style={
           rectangle,
           rounded corners,
           draw=blue, thin,
           text width=5em,
           minimum height=2em,
	fill= blue!10,
           text centered},
    pil/.style={
           ->,
           thick,
           shorten <=2pt,
           shorten >=2pt,}
}

\usepackage{pgfplots}
\usepackage{tikzscale}

\usepackage{algorithm}
\usepackage[noend]{algpseudocode}

\begin{document}

\title{\plaintitle}

\numberofauthors{2}
\author{%
  \alignauthor{Arun Kumar\\
   \affaddr{Dept. of Computer Science}\\
    \affaddr{University of Minnesota Twin Cities}\\
    \email{kumar250@umn.edu}}\\
  \alignauthor{Paul Schrater\\
   \affaddr{Depts. of Psychology and Computer Science}\\
    \affaddr{University of Minnesota Twin Cities}\\
    \email{schrater@umn.edu}}\\
}

\maketitle

\begin{abstract}
The vast majority of recommender systems model preferences as static or slowly changing due to observable user experience. However, spontaneous changes in user preferences are ubiquitous in many domains like media consumption and key factors that drive changes in preferences are not directly observable. These latent sources of preference change pose new challenges. When systems do not track and adapt to users' tastes, users lose confidence and trust, increasing the risk of user churn. We meet these challenges by developing a model of novelty preferences that learns and tracks latent user tastes. We combine three innovations: a new measure of item similarity based on patterns of consumption co-occurrence; model for {\em spontaneous} changes in preferences; and a learning agent that tracks each user's dynamic preferences and learns individualized policies for variety. The resulting framework adaptively provides users with novelty tailored to their preferences for change per se.
\end{abstract}



\section{Introduction}

An exponential growth in online retail and content delivery businesses like Amazon, Spotify, Netflix etc. provided an influx of choices to consumers. The amount of available information at our disposal clearly exceeds the search and consumption capacity of an individual. It creates two conflicting problems - an overabundance of choices but also a saturation of frequent yet not preferred content. In such situations, recommender systems play a crucial role in filtering suitable content for the user. However, recommender systems normally treat preferences as being fixed, an assumption profoundly violated in key recommender domains.  Rather than fixed, a user's affinity towards content can change with consumption and time
, repetitive or stale content leads to satiation, boredom or devaluation\cite{kapoor2013measuring}, and users are also intrinsically inclined to explore\cite{Kapoor2015}.

As a result, in the majority of the current recommender systems, after an initial period of satisfaction, users become increasingly bored of stale and repeated content that do not match their tastes. Dissatisfaction and distrust with the system are the consequence, because recommendations act similarly to \emph{advice} -- they are ignored or are greeted with hostility if the system's recommendations show a poor understanding of our tastes and cannot be trusted. Recommenders need to be more like a close aide who knows one's preferences well, whose suggestions are sensitive to our current tastes and to changes in our preferences. For example, phrases like ``I am bored of my play list'', are common in online music listening. Users start exploring on their own, indicating they have lost trust in the recommender system. 

\begin{figure}[h!]
\centering
\begin{tikzpicture}
[ auto,
    block/.style={rectangle, draw=blue, thick, fill=blue!20, text width=5em, align=center, rounded corners, minimum height=2em},
    block1/.style={ rectangle, draw=blue, thick, fill=blue!20, text width=5em, align=center, rounded corners, minimum height=2em },
    line/.style={draw,thick, -latex', shorten >=2pt},
    cloud/.style={ draw=red, thick, ellipse, fill=red!20,minimum height=1em}
 ]
\node[font=\scriptsize] at (0,5) (music){Music};
\node [below = 1pt of music, rectangle, draw, text width=4pt, text centered, minimum width=10pt, minimum height=12pt,font=\tiny] (S1) {S1};
\node [right = 1pt of S1,rectangle, draw, text width=4pt, text centered, minimum width=10pt, minimum height=12pt,font=\tiny] (S2) {S2};
\node [right = 1pt of S2,rectangle, draw, text width=4pt, text centered, minimum width=10pt, minimum height=12pt,font=\tiny](S3) {S3};
\node [right = 1pt of S3,rectangle, draw, text width=4pt, text centered, minimum width=10pt, minimum height=12pt,font=\tiny](S4) {$\hdots$};
\node [right = 1pt of S4,rectangle, draw, text width=4pt, text centered, minimum width=10pt, minimum height=12pt,font=\tiny](Sk) {Sk};

\path[->]  ($(S1.south)+(0,-0.1)$)  edge node {} ($(Sk.south)+(0,-0.1)$);
\node [font=\scriptsize] at ($(Sk.south)+(0.1,-0.15)$)  {t};

\node[font=\scriptsize] at (0,4) (video){Video};
\node [below = 1pt of video, rectangle, draw, text width=4pt, text centered, minimum width=10pt, minimum height=12pt,font=\tiny] (V1) {V1};
\node [right = 1pt of V1,rectangle, draw, text width=4pt, text centered, minimum width=10pt, minimum height=12pt,font=\tiny] (V2) {V2};
\node [right = 1pt of V2,rectangle, draw, text width=4pt, text centered, minimum width=10pt, minimum height=12pt,font=\tiny](V3) {V3};
\node [right = 1pt of V3,rectangle, draw, text width=4pt, text centered, minimum width=10pt, minimum height=12pt,font=\tiny](V4) {$\hdots$};
\node [right = 1pt of V4,rectangle, draw, text width=4pt, text centered, minimum width=10pt, minimum height=12pt,font=\tiny](Vl) {Vl};

\path[->]  ($(V1.south)+(0,-0.1)$)  edge node {} ($(Vl.south)+(0,-0.1)$);
\node [font=\scriptsize] at ($(Vl.south)+(0.1,-0.15)$)  {t};

\node [below = 1pt of V3,rectangle, text width=4pt, text centered, minimum width=10pt, minimum height=10pt,font=\small](D3) {$\vdots$};

\node[font=\scriptsize] at (0,2.5) (articles){Articles};
\node [below = 1pt of articles, rectangle, draw, text width=4pt, text centered, minimum width=10pt, minimum height=12pt,font=\tiny] (A1) {A1};
\node [right = 1pt of A1,rectangle, draw, text width=4pt, text centered, minimum width=10pt, minimum height=12pt,font=\tiny] (A2) {A2};
\node [right = 1pt of A2,rectangle, draw, text width=4pt, text centered, minimum width=10pt, minimum height=12pt,font=\tiny](A3) {A3};
\node [right = 1pt of A3,rectangle, draw, text width=4pt, text centered, minimum width=10pt, minimum height=12pt,font=\tiny](A4) {$\hdots$};
\node [right = 1pt of A4,rectangle, draw, text width=4pt, text centered, minimum width=10pt, minimum height=12pt,font=\tiny](Am) {Am};

\path[->]  ($(A1.south)+(0,-0.1)$)  edge node {} ($(Am.south)+(0,-0.1)$);
\node [font=\scriptsize] at ($(Am.south)+(0.1,-0.15)$)  {t};

\draw[dotted] ($(music.north west)+(-0.04,0.0)$)  rectangle ($(articles.south east)+(1.5,-0.7)$) node[align=center, below = 5pt of A3,font=\small, scale=0.9,text width=3cm] (content){Content \\ Consumption History};

\node at ($(content.north east)+(2.25,3.75)$) [rectangle, draw, text width=4em, text centered, font=\small, scale=0.9] (items){item-1\\item-2\\$\vdots$\\item-k};

\node at ($(content.north east)+(2,1.5)$) (problem) {
\input{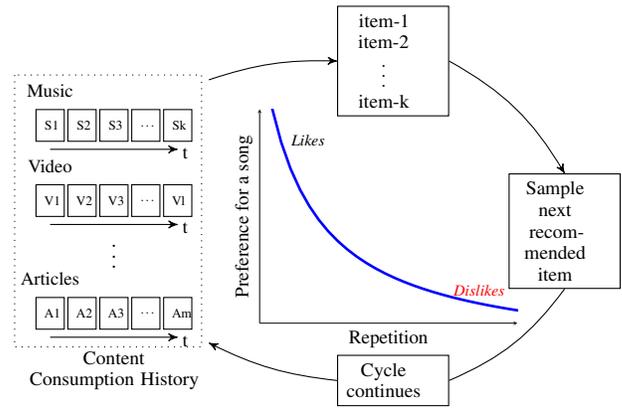}};

\node at ($(problem.north east)+(0.5,-1.75)$) [rectangle, draw, text width=4em, minimum height=8pt, minimum width=8pt, text centered, font=\small, scale=0.9] (recommend){Sample next recommended item};
\node at ($(content.north east)+(2.25,-0.5)$) [rectangle, draw, text width=4em, text centered, font=\small, scale=0.9] (cycle){Cycle continues};

\path[->]  ($(content.north east)+(-0.2,3.5)$)  edge  [bend left=10]  node {} (items.west);
\path[->]  (items.east)  edge [bend left=15] node {} (recommend.north);
\path  (recommend.south)  edge [bend left=15] node {} (cycle.east);
\path[->]  (cycle.west)  edge  [bend left=10] node {} ($(content.north east)+(-0.2,0.0)$);

\end{tikzpicture}
\caption{Illustration of the problems spontaneous preference changes create for recommender systems. Under repeated use, systems generate ranked itemized recommendations through a cycle by analyzing consumption histories of users in the system, then updates histories after user consumption. Sooner or later, the recommended item sets become repetitive, and increasingly risk becoming stale and boring to users, due to their spontaneous devaluation of repeated content, shown in the inset graph.}
\end{figure}

Figure 1 illustrates the problem. Recommender systems generate suggestions by analyzing the collection of consumption histories. The longer users interact with the system, the more likely recommendations will be dominated by a few items that have been frequently consumed, collapsing diversity and novelty in the system. Because items typical devalue with repeated exposure, the system will increasingly generate undesirable recommendations over time. As a result, current systems face two main challenges - tracking the changes in user preferences, and the absence of historical information to recommend new items to a user in the system. Without understanding the current state of users' tastes, it is hard for computational systems to match recommendations to users' liking. 

In psychology, changes in human preferences due to boredom or a desire for novelty are well known and characterized\cite{geiwitz1966structure, McAlister1982, vodanovich2003psychometric}, but little has been done to incorporate them into recommender systems. Boredom and a desire for novelty have been demonstrated in music listening\cite{kapoor2013measuring,Kapoor2015}, however, current systems do not have the means to understand or incorporate these latent preferences, largely due to the challenges of estimating and tracking these cognitive constructs and in using them to improve user's experiences. This creates a fundamental question: how can recommender systems understand when a user is bored or seeking novelty? Another question is how do we determine which user to target for recommendation of new items that do not have any prior history. We answer these two fundamental questions. 

To answer these questions on changing preferences with exposure and novel content, we introduce a novel, flexible and modular learning framework to represent user taste, track changes in user taste and learn the individualized policies matching user preferences, termed as \emph{Novelty Learning}. We treat preferences as {\em dynamically varying consumption bundles}.  Items are bundled together so that users prefer items in the bundle to recur in temporal proximity, but we allow the composition of these bundles to change over time. This model of preferences is novel, and we term it {\em proximity filtering}, because it relies solely on the consumption proximity between items in user's histories.  We also find a set of latent musical tastes that are shared across users. We track their dynamic preferences in taste space, which allows us to recommend both novel items to users based on their current taste preferences, and recommend new tastes as the current taste becomes stale. The resulting system is best termed novelty learning via collaborative proximity filtering. 

The goal of this paper is to introduce the novel concept of proximity filtering and demonstrate that this method can learn users' individualized novelty preferences in taste space. The remainder of this document presents this system in detail. Some of the common terms used are defined in table I. Section 2 provides background. In section 3 and 4, we discuss the presented model in details. The experimental set up is discussed in section 5, followed by discussion on results in the context in section 6. Finally, we conclude with a note on future scope.
\begin{table}[h]
\centering
\begin{tabular}{|p{1.75 cm}|p{6.1 cm}|}
\hline
\bf{Term} & \bf{ Meaning} \\ \hline
Taste & Latent theme discovered from items co-consumed - probabilistic playlists \\ \hline
Taste Profile & User profile based on discovered themes\\ \hline
Preferences & Taste bundles which are dynamic, hence dynamic preferences \\ \hline
Novelty Seeking & User preferences when a user consumes items with changing taste profile \\ \hline
Familiarity Content & User preferences when a user consumes items without a change in taste profile \\ \hline
Proximity Filtering & Similarity by proximity in consumption bundles \\ \hline
NoveltyLearn &  Novelty learning agent\\ \hline
Session & A time window of user's music listening activity\\
\hline
\end{tabular}
\newline
\caption{Terminology} 
\end{table}

\section{Background}
With most of the media being consumed in digital form now, online suppliers must target content to customer's preferences.  
Characterizing user preferences in domains like music that have vast and diverse content poses significant challenges. The sheer diversity requires aggregating content using similarities. 

Current recommender systems are driven by inducing similarity: either between users \cite{sarwar2001item}, deriving recommendations from other users with the most similar consumption history; or similarity between items. For music recommendations there have been many attempts to exploit both user and item similarity to predict preferences.  Oord et.al. \cite{NIPS2013-5004} propose a latent factor model using weighted matrix factorization and deep convolutional neural networks. They show that recommendations using additional attributes are more sensible and useful when usage data is sparse. Fabian et.al.\cite{abel2013cross} evaluate tag based user profiles from social media in a personalization effort. Their work focuses on cross-system user modeling to improve recommendation quality. These methods are powerful, but as illustrated in the left panel of Figure 2, this induced similarity relies critically on \emph{static} preferences, while users' preferential tastes are inherently more dynamic . 

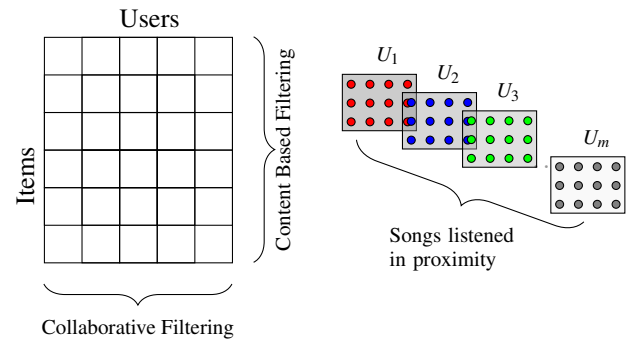
\begin{figure}[h!]
\centering
\begin{tikzpicture}
    [ box/.style={rectangle,draw=black,thin, minimum size=1cm}, scale = 0.5]
\foreach \x in {0,1,...,3}{
    \foreach \y in {0,1,...,4}
        \node[box] at (\x,\y) (box){};
}
\node [text width = 3.5cm] at ($(box.north)+(1.5,0.5)$) (users) {Users};
\node [text width = 3.5cm, rotate=90] at ($(users.south)+(-6.0,-1.0)$) (items) {Items};

\draw[decorate,decoration={brace,amplitude=8pt,raise=8pt}] ($(box.north)+(1.0,0)$) -- ($(box.north)+(1.0,-6.0)$) node (content) [black,midway,xshift=+20pt,yshift=+0pt,rotate=90] {\small{Content Based Filtering}};

\draw[decorate,decoration={brace,amplitude=8pt,raise=8pt,mirror}] ($(box.west)+(-3.0,-5.0)$) -- ($(box.west)+(2.0,-5.0)$) node (collaborate) [black,midway,xshift=0pt,yshift=-25pt,rotate=0] {\small{Collaborative Filtering}};

\node [text width = 4.5cm,font=\small] at ($(content.north)+(6,+2.5)$) (nohistory) {};
    \node (asr1)  at ($(nohistory.south)+(-3.0,-1.0)$) [draw, fill=gray!100, text width=0.75cm, text centered, minimum height=0.75cm, fill opacity=0.4]{};
    \node[above,font=\small] at ($(asr1.north)+(0.25,0)$) {$U_1$};
    \node (asr2) at ($(asr1.north east)+(0.6,-1.25)$)  [draw, fill=gray!80, text width=0.75cm, text centered, minimum height=0.75cm, fill opacity=0.4] {};
    \node[above,font=\small] at ($(asr2.north)+(0.25,0)$) {$U_2$};
    \node (asr3) at ($(asr2.north east)+(0.6,-1.25)$)  [draw, fill=gray!60, text width=0.75cm, text centered, minimum height=0.75cm, fill opacity=0.4] {};
    \node[above,font=\small] at ($(asr3.north)+(0.25,0)$) {$U_3$};
    \node (dots4) at ($(asr3.north east)+(0.25,-1.5)$)  [text width=0.75cm, text centered, minimum height=0.75cm, fill opacity=0.4] {$\hdots$}; 
    \node (asr4) at ($(dots4.north east)+(0.1,-1.25)$)  [draw, fill=gray!10, text width=0.75cm, text centered, minimum height=0.75cm, fill opacity=0.4] {};
    \node[above,font=\small] at ($(asr4.north)+(0.25,0)$) {$U_m$};

\foreach \x in {0,0.5,...,1.5}{
    \foreach \y in {0,0.5,...,1.0}
        \node[draw,circle,inner sep=0pt,minimum size=3pt,fill=red] at ($(asr1.south)+(\x-0.75,\y+0.25)$) (songs1){};
}

\foreach \x in {0,0.5,...,1.5}{
    \foreach \y in {0,0.5,...,1.0}
        \node[draw,circle,inner sep=0pt,minimum size=3pt,fill=blue] at ($(asr2.south)+(\x-0.75,\y+0.25)$) (songs2){};
}

\foreach \x in {0,0.5,...,1.5}{
    \foreach \y in {0,0.5,...,1.0}
        \node[draw,circle,inner sep=0pt,minimum size=3pt,fill=green] at ($(asr3.south)+(\x-0.75,\y+0.25)$) (songs3){};
}

\foreach \x in {0,0.5,...,1.5}{
    \foreach \y in {0,0.5,...,1.0}
        \node[draw,circle,inner sep=0pt,minimum size=3pt,fill=gray] at ($(asr4.south)+(\x-0.75,\y+0.25)$) (songs4){};
}

\draw[decorate,decoration={brace,amplitude=10pt,raise=15pt,mirror}] ($(asr1)+(-0.25,0)$) -- ($(asr4)+(+0.25,0)$) node [black,midway,xshift=0pt,yshift=-40pt,text width = 2.5cm,font=\small] {Songs listened \\ in proximity};


\end{tikzpicture}
\caption{{\em Left Panel} Collaborative filtering and content based filtering both rely solely on data in user/item matrices that have no explicit representation of time. User/item pairs are usually sparsely sampled and thus change slowly as new user/item choices enter the system.  Attribute similarities in content-based filtering are normally predefined and fixed. The result is that both methods implicitly assume near static preferences, creating difficulties in tracking more rapidly varying dynamic preferences.  {\em Right Panel} Collaborative Proximity Filtering instead relies on finding shared proximity structure across user's song histories. It can be viewed as finding an implicit User/Taste matrix, where tastes are defined by sets of songs listened in close proximity (latent playlists). Recent song choices can be used to track and forecast preference dynamics in taste space.  }
\end{figure}

Moore et.al.\cite{moore2013taste} analyze listening preferences of a population of users to try to track a hypothesized gradual change in group preferences over time,  ignoring individual tastes. To personalize recommendations, Yin et.al.\cite{yin2014temporal} explore social scenarios and study the influence of other users' choices and ratings on user interests. The rating mechanism, though explicit, is problematic because it is either a delayed reflection of user's state of mind when item was consumed or it can become influenced by other user's choices and public attention. This work  also assumes that users' intrinsic interests do not vary much and that temporal dynamics comes from changes in public attention. 

In domains like music, users' preferences clearly change individually with time and and should be treated as intrinsically dynamic. This means that all interest can be lost in items that were previously favorites (over-exposure breeds contempt\cite{givon1984variety} \cite{jeuland1979brand}). Users also have well-characterized desire for novel content \cite{berlyne1960conflict}, for which the recommender system will have little or no data. Rapid preference changes like these can catastrophically degrade the user's experience with the recommender system, unless it adapts and tracks these changes. Moreover, different users have different preferences dynamics, and our recommendations also should be adapted to their individualized appetites to explore new options.

In music listening, users have a natural preference dynamics that has previously been identified \cite{kapoor2013measuring}. Preference for items increases with initial exposure, while boredom (item devaluation) sets in with subsequent exposures. Preferences tend to partially recover with time away from an item. Users recurrently consume a small, working set of items over days and weeks, slowly removing old songs and introducing new ones (essentially an evolving playlist).  Viewing a window of listening history as a bundle of items co-consumed, we can say that listening preferences dynamically shift between consumption bundles of songs we term {\em tastes}. Similarity between songs is induced by being part of the same consumption bundle: similarity via temporal proximity. To better track changes in user's preferences, we exploit this recurrent structure in music listening, using song proximity to filter collaborative and novel recommendations to users (see Figure 2, right panel).    

\section{Proximity Filtering by Tastes Discovery}
Users consume media items like songs in temporal proximity of consumption bundles and not in isolation. These consumption bundles are meaningful in understanding user's preferences at an abstract level. The items within the same consumption bundle are likely to be similar in user's tastes and they are likely to be co-preferred because of the proximity in the consumption bundles as illustrated in Figure 3. It forms the basis of our intuition of proximity filtering.
\begin{figure}[h!]
\centering
\begin{tikzpicture}[auto, scale =0.6, show background rectangle]
\node[rectangle,draw,text width=4em, text centered, minimum height=1.5em, fill=white] (user) {User};
\node[rectangle,draw,text width=4em, text centered, minimum height=1.5em, right=7em of user,fill=white] (history) {History};
    \path +(5.5,-3.0) node (asr1) [draw, fill=blue!100, text width=1.5em, text centered, minimum height=1.5em,drop shadow, fill opacity=0.4] {};
    \path +(6.3,-2.75) node (asr2)[draw, fill=blue!80, text width=1.5em, text centered, minimum height=1.5em,drop shadow, fill opacity=0.4] {};
    \path +(7.1,-2.5) node (asr3)[draw, fill=blue!60, text width=1.5em, text centered, minimum height=1.5em,drop shadow, fill opacity=0.4] {};
    \path +(8.2,-2.45) node (dots)[above, text width=3em, text centered] {$\hdots$}; 
    \path +(9.2,-2.0) node (asr4)[draw, fill=blue!10, text width=1.5em, text centered, minimum height=1.5em,drop shadow, fill opacity=0.4] {};
\node (l1) at (5.9,-3.0) [draw,circle,inner sep=0pt,minimum size=6pt,fill=red] {};
\node (l2) at (6.8,-3.0) [draw,circle,inner sep=0pt,minimum size=6pt,fill=red] {};

\draw[decorate,decoration={brace,amplitude=8pt,raise=8pt}] (asr1) -- (asr4) node [black,midway,xshift=+50pt,yshift=+15pt] {Consumption Bundles};
\node[below=4em of user, draw,ellipse,minimum height=3em, minimum width=6em, align=center,fill={rgb:red,1;green,2;blue,4},text=white,font=\bf] (proximity) at (0,0) {Proximity \\ Filtering};

\path[->]  (user.south)  edge   [bend right=45]   node {} (proximity);
\path[->]  (history.south)  edge   [bend right=15]   node {} (proximity);

\node[text width=12em, below=of asr1,align=center] at (8.0,-2.0) (text1){Similarity by proximity; Likely to be co-preferred};
   \path[->]  (l1.south)  edge  node {} (text1.north);
   \path[->]  (l2.south)  edge  node {} (text1.north);

\end{tikzpicture}
\caption{Proximity filtering - items like songs consumed in proximity are likely to be co-preferred}
\end{figure}
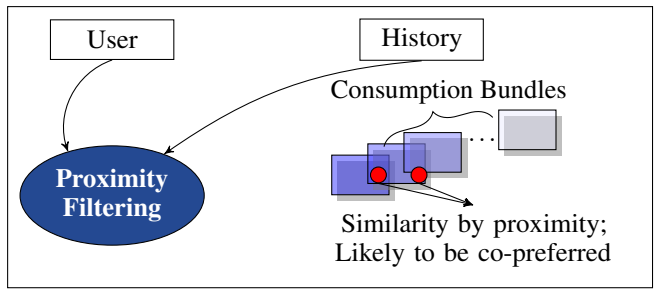
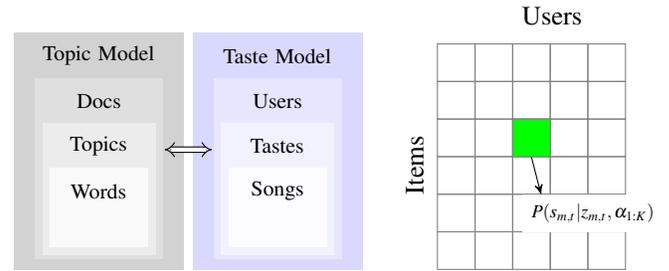
\begin{figure}[h!]
\centering
\begin{tabular}{cc}
\begin{tikzpicture}
[ auto,
    block/.style={rectangle, draw=blue, thick, fill=blue!20, text width=5em, align=center, rounded corners, minimum height=2em},
    block1/.style={ rectangle, draw=blue, thick, fill=blue!20, text width=5em, align=center, rounded corners, minimum height=2em },
    line/.style={draw,thick, -latex', shorten >=2pt},
    cloud/.style={ draw=red, thick, ellipse, fill=red!20,minimum height=1em},
    outer sep=0,inner sep=0
 ]

\node[rectangle, align=center, text width=1.75cm, minimum width=0.5cm, minimum height=0.5cm, font=\small] at (0,0.0) (block1){Topic Model};
\node [below = 0.1cm of block1, rectangle, align=center, text width=1cm, minimum width=1cm, minimum height=0.5cm, font=\small] (block11) {Docs};
\node [below = 0.1cm of block11, rectangle, align=center, text width=1cm, minimum width=1cm, minimum height=0.5cm, font=\small] (block12) {Topics};
\node [below = 0.1cm of block12, rectangle, align=center, text width=1cm, minimum width=1cm, minimum height=0.5cm, font=\small] (block13) {Words};

\begin{pgfonlayer}{background}
	\fill[gray!35] ([shift={(-0.25,0.05)}]block1.north west) rectangle ([shift={(0.25,-2.6)}]block1.south east);
	\fill[gray!25] ([shift={(-0.35,0.05)}]block11.north west) rectangle ([shift={(0.35,-1.85)}]block11.south east);
	\fill[gray!15] ([shift={(-0.25,0.05)}]block12.north west) rectangle ([shift={(0.25,-1.2)}]block12.south east);
	\fill[gray!05] ([shift={(-0.15,0.05)}]block13.north west) rectangle ([shift={(0.15,-0.5)}]block13.south east);
\end{pgfonlayer}

\node[rectangle, align=center, text width=1.75cm, minimum width=0.5cm, minimum height=0.5cm, font=\small] at ([shift={(1.5,0)}]block1.east) (block2){Taste Model};
\node [below = 0.1cm of block2, rectangle, align=center, text width=1cm, minimum width=1cm, minimum height=0.5cm, font=\small] (block21) {Users};
\node [below = 0.1cm of block21, rectangle, align=center, text width=1cm, minimum width=1cm, minimum height=0.5cm, font=\small] (block22) {Tastes};
\node [below = 0.1cm of block22, rectangle, align=center, text width=1cm, minimum width=1cm, minimum height=0.5cm, font=\small] (block23) {Songs};

\begin{pgfonlayer}{background}
	\fill[blue!15] ([shift={(-0.25,0.05)}]block2.north west) rectangle ([shift={(0.25,-2.6)}]block2.south east);
	\fill[blue!10] ([shift={(-0.35,0.05)}]block21.north west) rectangle ([shift={(0.35,-1.85)}]block21.south east);
	\fill[blue!05] ([shift={(-0.25,0.05)}]block22.north west) rectangle ([shift={(0.25,-1.2)}]block22.south east);
	\fill[blue!01] ([shift={(-0.15,0.05)}]block23.north west) rectangle ([shift={(0.15,-0.5)}]block23.south east);
\end{pgfonlayer}
\draw[implies-implies,double equal sign distance,] ([shift={(0,-1.25)}]block1.east) -- ([shift={(0,-1.25)}]block2.west);

\end{tikzpicture}
&
\begin{tikzpicture}
    [ box/.style={draw=gray,thin, minimum size=0.5cm}, scale = 1]
\node [] at ($(block2.north)+(0,4)$) (matrix) {};
\foreach \x in {0,0.5,...,2.0}{
    \foreach \y in {0,0.5,...,2.5}
        \node[box] at (\x,\y) (box){};
}
\node [text width = 2.0cm] at ($(matrix.north)+(-0.5,-1.25)$) (users) {Users};
\node [text width = 2.0cm, rotate=90] at ($(matrix.west)+(-2.8,-2.5)$) (items) {Items};
\node[box,fill=green] at (1,1.5) (boxcolor) {};
\node [rectangle,fill=white,text width = 2.5cm,font=\scriptsize] at ($(matrix.west)+(0,-3.75)$) (tag) {$P(s_{m,t}|z_{m,t},\alpha_{1:K})$};
\path[->]  (boxcolor.south)  edge  node {} ($(tag.north west) + (0.25,0)$);

\end{tikzpicture}
\end{tabular}
\caption{\emph{Left panel} Analogy with standard Latent Dirichlet Allocation: Taste Model treats users content consumption as documents. The discovered taste themes are analogous to topics in text documents and consumed songs are analogous to words. \emph{Right panel} Users consume items/songs according to their tastes. Each cell represents probability of a song in user's taste profile.}
\end{figure}
\begin{figure}[h!]
\centering
\begin{tikzpicture}
\node (beta) at (0,0) [circle, draw,label=-90:$\beta_{m}$] {};
\node (z) at ($(beta.east)+(1,0)$) [circle, draw,label=-90:$z_{m,n}$] {};
\node (s) at ($(z.east)+(1.5,0)$) [circle, draw, fill=gray, label=-90:$s_{m,n}$] {};
\node (alpha) at ($(s.east)+(2.0,0)$) [circle, draw, label=-90:$\alpha_{k}$] {};

\draw[->] (beta) -- (z);
\draw[->] (z) -- (s);
\draw[->] (alpha) -- (s);

\draw ($(z.south)+(-0.5,-0.75)$) rectangle ($(s.north)+(0.5,.5)$);
\draw ($(beta.south)+(-0.5,-1.25)$) rectangle ($(s.north)+(1,0.75)$);
\draw ($(alpha.north)+(-0.5,0.75)$) rectangle ($(alpha.south)+(0.5,-1.25)$);

\node [font=\scriptsize] (ntracks) at ($(s.south)+(-0.25,-0.55)$) {N Tracks};
\node [font=\scriptsize] at ($(ntracks.south)+(0.5,-0.25)$) {M Users};
\node [font=\scriptsize] at ($(alpha.south)+(0,-1)$) {K Tastes};
\end{tikzpicture}
\caption{Tastes Discovery: In user music listening history, the songs $s_{m,n}$ are observed. The discovered tastes are common to all the users and lie outside the users' activity space, in plate K tastes. Every user has a proportion z of the discovered tastes. Uncovering these latent tastes is key to the taste space representation.}
\end{figure}
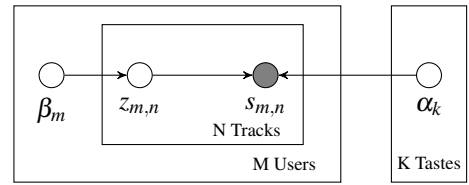

We introduce a novel computational proximity filtering model that considers user tastes as latent structure over dynamic consumption bundles. To discover the underlying common pool of users' tastes, we adapt topic discovery models used in text analysis to discover the equivalent musical topics from listening histories. Topic modeling methods discover useful themes from word co-occurrence frequencies within documents, by analyzing a large text corpus. We adapt Latent Dirichlet Allocation \cite{blei2003latent}, a powerful topic modeling technique through an analogy between the set of music listening histories and a text corpus.  Figure 4 illustrates this analogy between tastes discovery and a document topic model. Documents correspond to contiguous sections of listening history from a user.  The tastes correspond to latent topics in a standard topic model. All the users share the same set of discovered tastes but each user has a probabilistic distribution over these latent tastes. User preferences are dynamic, therefore, we represent a user's \emph{dynamic preference state} as a time-varying mixture of latent tastes. User behavior in content consumption domain is a manifestation of these dynamic preferences.

The discovered taste themes are not predefined but inherently data driven and provide a meaningful and interpretable representation of users' liking. In natural language processing, language modelling methods learn relationships between language constructs. Since our aim is to uncover latent themes to understand user behaviors and liking, topic modeling is an appropriate and preferred method.  

More specifically, our tastes model maps into a topic model as illustrated in Figure 5. Each user $U_{m}$ consumes content $s_{m,n}$ and we can represent the observed content as derived from latent tastes. Each of the user tastes is one among the common pool of the discovered tastes and derived from a taste proportion. Figure 4 illustrates a matrix form with each cell representing the probability of a song in user's taste proportion and the taste, in turn, is derived from the discovered common pool of latent tastes. Similar to \cite{blei2003latent}, each cell in this matrix represents
\begin{equation}
P(s_{m,n} | z_{m,n},\alpha_{1:K}) = \alpha_{z_{m,n},s_{m,n}} 
\end{equation}
Discovered tastes as well as taste proportions for users are assumed to come from a Dirichlet. Therefore, the model is reduced to finding a joint distribution similar to \cite{blei2003latent}
\begin{multline}
P(\alpha_{1:K}, \beta_{1:M},z_{1:M},s_{1:M}) = \Pi_{i=1}^K P(\alpha_i) \Pi_{m=1}^M P(\beta_m) \\
(\Pi_{n=1}^N P(z_{m,n}|\beta_m) P(s_{m,n}|z_{m,n},\alpha_{1:K})) 
\end{multline}
\begin{figure*}[htp!]
\centering
\input{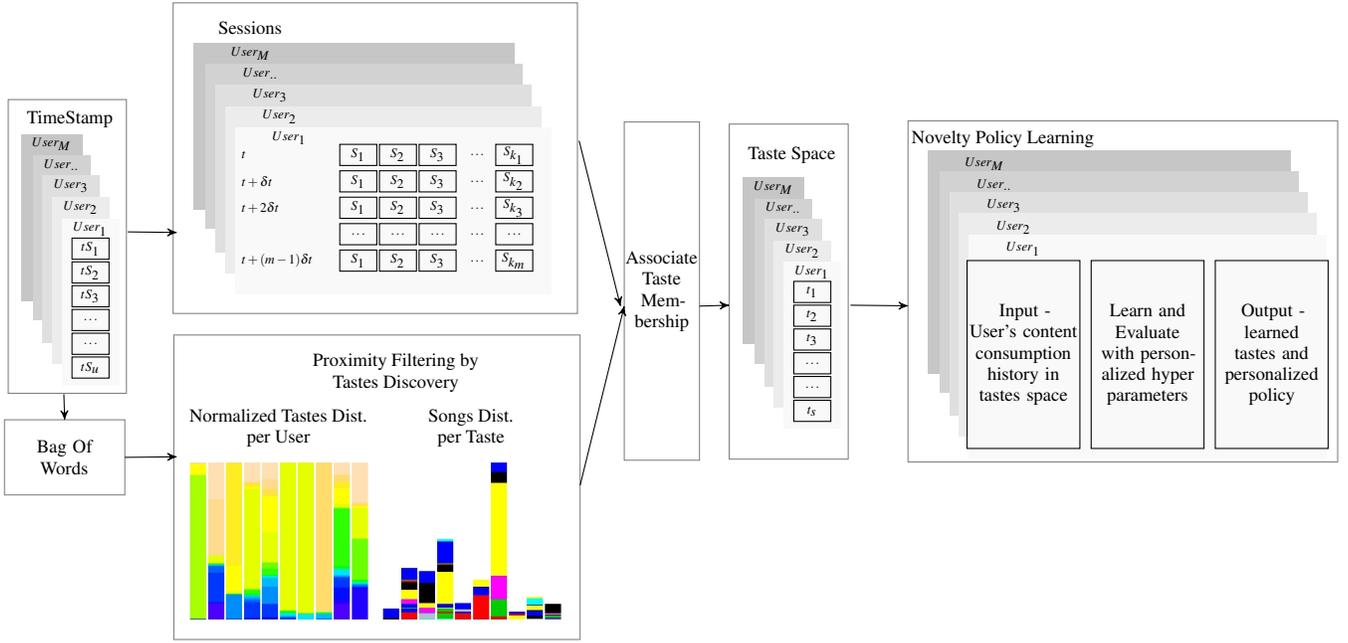}
\caption{Model: Users consume items at a granular level but their liking or preferences are abstract and implicit. We introduce proximity filtering by discovering the latent tastes. The itemized space is transformed into a taste space to provision for learning users' novelty seeking behaviors.}
\end{figure*}
Until recently, users were largely profiled using static attributes like gender, location etc., but such static attributes are not of much value \cite{wired-netflix}. We uncover user attributes by understanding the distinctive mixture of their latent tastes as well as the similarities in a common taste pool. This interpretation lies at the core of taste space transformation to learn \emph{inherent} user preferences and novelty seeking behaviors.

Our model consists of modular building blocks providing flexibility, re-usability and extensibility to a range of content delivery domains. We present the schematic diagram of the framework in Figure 6. In the taste discovery, we obtain tastes distribution for users and tracks distribution for tastes. The tracks distribution for tastes demonstrate that different tastes vary in proportions of tracks. As we observe, the tastes have differentiating proportions of a set of tracks and these proportional compositions characterize the discovered tastes. While the discovered tastes are shared across users, users have a distinctive proportional mixture of latent tastes. This representation allows us to transform users' itemized consumption into latent taste space using the discovered tastes pool and learn users' novelty seeking behaviors. In a nutshell, the discovered latent tastes are leveraged to learn user preferences and we demonstrate our method for users' online music listening.
\section{Novelty Learning}
In Psychology, boredom is often considered as an emotion of dullness or lacking stimulation. Goetz et.al. \cite{Goetz2013} describe and support the external validity of boredom with varying degrees of arousal and valence. However, due to its hidden presentation in perceived environment, it is hard to track boredom using explicitly observable signals like ratings of items, satisfaction scores etc. In an academic setting, Nett et al. \cite{Nett2010626} explored different strategies with groups of individuals to find out their activities in boredom situations. In absence of any visual or observable cue about user's emotional state, recommender systems fail to understand whether a user is novelty seeking or not and users, in absence of content matching their taste, get bored sooner than later.
\begin{figure}[h!]
\centering
\includegraphics[width=0.8\linewidth,height=3.5cm]{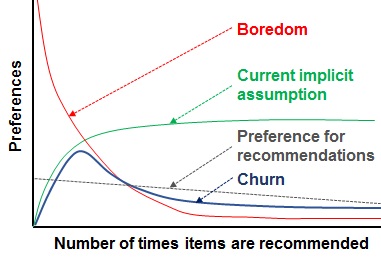}
\caption{Devaluation of user preferences for recommended items implies that the boredom settles in and user might not trust the system any longer. From systems' point of view, without any proactive intervention to improve user satisfaction, this user might stop using the provided service altogether. Therefore, a proactive mechanism to learn which users are seeking novelty proactively is critical.}
\end{figure}

Figure 7 illustrates devaluation of user preferences as user receives recommendations. In the beginning, users might have a stronger liking to the recommended items, however, as the cycle of repetitive or stale recommendations continues, users get bored and stop trusting the system leading to either user churn or disengagement. Boredom increases with repeated exposure. Since boredom does not manifest itself as visibly as other emotional states like anger or excitement but a more silent one, novelty seeking users might not even provide a feedback or any directly observable signal. It makes current recommender systems completely clueless to understand users' tastes. Once a user is disengaged, it is an uphill task for the systems to earn user's trust back. Therefore, we need a machinery to understand user behaviors more proactively and we aim to bridge this fundamental gap by providing a method to learn novelty seeking behavior using latent tastes. 

\subsection{Taste Membership}
Users consume contents at items level like songs but their implicit behaviors and preferences can not be understood in item level view but an abstract view. The items are consumed in bundles which reflect users' affinity to a taste. Such a representation is useful in understanding implicit constructs that drive preferences.Therefore, we use the discovered tastes pool to associate each of the item bundles with a taste and transform temporal sequences of users' consumed items bundles into an abstracted taste space. Figure 8 illustrates the method.
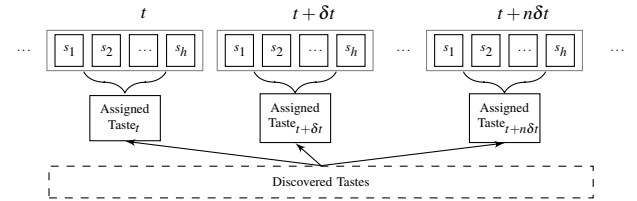
\begin{figure}[h!]
\centering
\begin{tikzpicture}[auto, node distance=2cm,>=latex']

\node at (1,0) [font=\scriptsize] (b0){};
\node at ($(b0) + (-1,-0.5)$) [rectangle, text width=4pt, text centered, minimum width=10pt, minimum height=12pt,font=\tiny] (S01) {$\hdots$};

\node at ($(b0.east)+(0.5,0)$) [font=\scriptsize] (b1){$t$};
\node at ($(b1) + (-1,-0.5)$) [rectangle, draw, text width=4pt, text centered, minimum width=10pt, minimum height=12pt,font=\tiny] (S11) {$s_1$};
\node [right = 3pt of S11,rectangle, draw, text width=4pt, text centered, minimum width=10pt, minimum height=12pt,font=\tiny] (S12) {$s_2$};
\node [right = 3pt of S12,rectangle, draw, text width=4pt, text centered, minimum width=10pt, minimum height=12pt,font=\tiny](S13) {$\hdots$};
\node [right = 3pt of S13,rectangle, draw, text width=4pt, text centered, minimum width=10pt, minimum height=12pt,font=\tiny](S1h) {$s_h$};
\draw[decorate,decoration={brace,amplitude=8pt,raise=8pt,mirror}] (S11) -- (S1h) node (recency) [black,midway,xshift=+0pt,yshift=+0pt] {};
\node at ($(S12) + (0.25,-0.9)$) [rectangle, draw, text width=20pt, text centered, minimum width=10pt, minimum height=12pt,font=\tiny] (at1) {Assigned Taste$_{t}$};
\draw [gray] ([shift={(-0.1,0.05)}]S11.north west) rectangle ([shift={(0.1,-0.05)}]S1h.south east);  
\node at ($(b1) + (2.35,-2.25)$) [rectangle, draw, dashed, text width=7cm, text centered, minimum width=1cm, minimum height=12pt,font=\tiny] (tastes) {Discovered Tastes};
\path[->] ($(tastes.north)+(0,0)$)  edge node {} ($(at1.south)+(0,0)$);

\node at ($(b1.east)+(2.1,0)$) [font=\scriptsize] (b2){$t+\delta t$};
\node at ($(b2) + (-1,-0.5)$) [rectangle, draw, text width=4pt, text centered, minimum width=10pt, minimum height=12pt,font=\tiny] (S21) {$s_1$};
\node [right = 3pt of S21,rectangle, draw, text width=4pt, text centered, minimum width=10pt, minimum height=12pt,font=\tiny] (S22) {$s_2$};
\node [right = 3pt of S22,rectangle, draw, text width=4pt, text centered, minimum width=10pt, minimum height=12pt,font=\tiny](S23) {$\hdots$};
\node [right = 3pt of S23,rectangle, draw, text width=4pt, text centered, minimum width=10pt, minimum height=12pt,font=\tiny](S2h) {$s_h$};
\draw[decorate,decoration={brace,amplitude=8pt,raise=8pt,mirror}] (S21) -- (S2h) node (recency) [black,midway,xshift=+0pt,yshift=+0pt] {};
\node at ($(S22) + (0.25,-0.9)$) [rectangle, draw, text width=20pt, text centered, minimum width=10pt, minimum height=12pt,font=\tiny] (at2) {Assigned Taste$_{t+\delta t}$};
\draw [gray] ([shift={(-0.1,0.05)}]S21.north west) rectangle ([shift={(0.1,-0.05)}]S2h.south east);  
\path[->] ($(tastes.north)+(0,0)$)  edge node {} ($(at2.south)+(0,0)$);

\node at ($(b2.east)+(1.75,0)$) [font=\scriptsize] (b3){};
\node at ($(b3) + (-1,-0.5)$) [rectangle, text width=4pt, text centered, minimum width=10pt, minimum height=12pt,font=\tiny] (S31) {$\hdots$};

\node at ($(b3.east)+(0.5,0)$) [font=\scriptsize] (b4){$t+n\delta t$};
\node at ($(b4) + (-1,-0.5)$) [rectangle, draw, text width=4pt, text centered, minimum width=10pt, minimum height=12pt,font=\tiny] (S41) {$s_1$};
\node [right = 3pt of S41,rectangle, draw, text width=4pt, text centered, minimum width=10pt, minimum height=12pt,font=\tiny] (S42) {$s_2$};
\node [right = 3pt of S42,rectangle, draw, text width=4pt, text centered, minimum width=10pt, minimum height=12pt,font=\tiny](S43) {$\hdots$};
\node [right = 3pt of S43,rectangle, draw, text width=4pt, text centered, minimum width=10pt, minimum height=12pt,font=\tiny](S4h) {$s_h$};
\draw[decorate,decoration={brace,amplitude=8pt,raise=8pt,mirror}] (S41) -- (S4h) node (recency) [black,midway,xshift=+0pt,yshift=+0pt] {};
\node at ($(S42) + (0.25,-0.9)$) [rectangle, draw, text width=20pt, text centered, minimum width=10pt, minimum height=12pt,font=\tiny] (at4) {Assigned Taste$_{t+n\delta t}$};
\draw [gray] ([shift={(-0.1,0.05)}]S41.north west) rectangle ([shift={(0.1,-0.05)}]S4h.south east);  
\path[->] ($(tastes.north)+(0,0)$)  edge node {} ($(at4.south)+(0,0)$);

\node at ($(b4.east)+(1.75,0)$) [font=\scriptsize] (b5){};
\node at ($(b5) + (-1,-0.5)$) [rectangle, text width=4pt, text centered, minimum width=10pt, minimum height=12pt,font=\tiny] (S51) {$\hdots$};

\end{tikzpicture}
\caption{Taste Membership Assignment - temporal windows of consumed items are transformed into taste space using discovered tastes pool.}
\end{figure}
We compute taste association i.e. a taste to be assigned to a temporal window. Let $s_{i}$ represent song or track id, $k$, tastes and $h$ be the history of tracks listened in a time period of length $n$ given by [$s_{1},s_{2},...s_{n}$]. For every temporal window, we compute the probability that a taste is associated with it and for a history $h$, it is given by
\begin{equation}
P(k|h) = \Pi_{i=1}^{n} P(s_i|k) P(k) / \Sigma_{i=1}^{n} P(s_i|k)P(k)
\end{equation}
where $P(k)$ = user taste probability and $P(s_i|k)$ = taste track probability
This evaluated probability is used to assign a taste to each temporal window i.e. session as 
\begin{equation}
Taste_{assigned}= max_k P(k|h)
\end{equation}
Consequently, users' items consumption space is transformed into a latent taste space and we can now learn users' preferences.

\subsection{Novelty Policy Q-Learning}
Learning users' behaviors like novelty seeking is crucial but ignored aspect in not only recommender systems but also human factors and engagement initiatives. An implicit nature of boredom manifestation also introduces difficulty in computational tracking of user behavior. To answer this problem, we aim to learn the users' novelty seeking policies. Since item level consumption view does not present insights abstract enough to learn behaviors, we need to resort to a taste space representation. Users tastes are dynamic and we present changing tastes as novelty seeking behavior, it also represents boredom with content. When users continue to consume items without a change in their tastes, they are content with the familiarity. Our novelty learning agent, a model free reinforcement learning agent, learns users' novelty seeking policies.

Reinforcement learning \cite{sutton1998reinforcement}  has its roots in psychology and animals' reward seeking behavior. The goal is to maximize the utility by taking a set of actions in an environment state and learn the policy that maximizes the utility. The environment is typically conceptualized as a Markov Decision Process. The reinforcement learning methods differ from classic supervised learning in a way that we do not know the correct input and output pairs nor there is any attempt to make an explicit correction. Rather, the models are based on experiencing the reward and exploitation-exploration trade-offs to find out the optimal or maximizing utility. Such level of unknown input-output pairs make it suitable to our learning problem. 
\begin{figure}[h!]
\centering
\begin{tikzpicture}[auto, node distance=2cm,>=latex']
    \node [coordinate, name=input] {};
    \node [coordinate, node distance=1.0cm, right of=input, name=sum] {};
    \node [draw, fill=blue!20, rectangle, text width = 2.52cm, align=center, minimum height=8pt, minimum width=30pt, right of=sum,font=\scriptsize] (controller) {Q-Learning};
    \node [coordinate, right of=controller,node distance=2.5cm] (output) {};
    \node at ($(controller.south)+(0,-0.5)$) [draw, fill=blue!20, rectangle, text width = 2.42cm, align=center, minimum height=8pt, minimum width=20pt, font=\scriptsize] (measurements) {Taste Space};
    \draw [draw,-] (input) -- node [font=\scriptsize] {Initial s} (sum);
    \draw [->] (sum) -- node {} (controller);
    \draw [->] (controller) -- node [name=y,font=\scriptsize] {Policy $\pi $}(output);

\node[font=\scriptsize,below = 1pt of measurements] (ra){Reward};
\node at ($(ra) + (-1,-0.35)$) [rectangle, draw, text width=4pt, text centered, minimum width=10pt, minimum height=12pt,font=\tiny] (S1) {S1};
\node [right = 3pt of S1,rectangle, draw, text width=4pt, text centered, minimum width=10pt, minimum height=12pt,font=\tiny] (S2) {S2};
\node [right = 3pt of S2,rectangle, draw, text width=4pt, text centered, minimum width=10pt, minimum height=12pt,font=\tiny](S3) {S3};
\node [right = 3pt of S3,rectangle, draw, text width=4pt, text centered, minimum width=10pt, minimum height=12pt,font=\tiny](S4) {$\hdots$};
\node [right = 3pt of S4,rectangle, draw, text width=4pt, text centered, minimum width=10pt, minimum height=12pt,font=\tiny](Sh) {Sh};

\draw[decorate,decoration={brace,amplitude=8pt,raise=8pt,mirror}] (S1) -- (Sh) node (recency) [black,midway,xshift=+50pt,yshift=+15pt] {};

\node [below = 17pt of Sh,rectangle, draw, text width=4pt, text centered, minimum width=10pt, minimum height=12pt,font=\tiny] (Sk) {Sk};
\node at ($(S3.south)+(0.15,-0.83)$) [rectangle, draw, text width=16pt, text centered, minimum width=10pt, minimum height=12pt,font=\tiny] (Seval) {Evaluate change in taste};
\node at ($(Seval.west)+(-0.6,0)$) [rectangle, draw, text width=16pt, text centered, minimum width=10pt, minimum height=12pt,font=\tiny] (reward) {Calculate reward};
\path[->]  ($(Sk.west)+(0,0)$)  edge node {} ($(Seval.east)+(0,0)$);
\path[->]  ($(Seval.west)+(0,0)$)  edge node {} ($(reward.east)+(0,0)$);

\draw [gray] ([shift={(-0.13,0.3)}]S1.north west) rectangle ([shift={(0.15,-0.35)}]Sk.south east);  
\draw [gray] ([shift={(-0.075,0.1)}]measurements.north west) rectangle ([shift={(0.25,-0.4)}]Sk.south east);  
\node[below = 1pt of measurements] (tastespace){};

\draw [->,font=\scriptsize] (y) |-  node[name=u] {} ($(tastespace.east) + (1.25,-0.75)$);
\node [font=\scriptsize] at ($(tastespace.east) + (2.5,0.2)$) {action a};
\draw [draw,->] ($(tastespace.west) + (-1.3,-0.6) $) -- ($(tastespace.west) + (-1.8,-0.6) $) -- 
($(tastespace.west) + (-1.8,0.95) $) -- ($(controller)+(-1.35,-0.15)$);
\draw [draw,->] ($(tastespace.west) + (-1.3,-0.8) $) -- ($(tastespace.west) + (-2.0,-0.8) $) -- 
($(tastespace.west) + (-2.0,1.05) $) -- ($(controller)+(-1.35,-0.07)$);
\node [] at ($(tastespace.west) + (-1.6,+0.1) $) {$r$};
\node [] at ($(tastespace.west) + (-2.2,+0.15) $) {$s'$};

\end{tikzpicture}
\caption{Novelty learning agent is a model-free Q-learning agent in taste space. It learns an optimal policy $\pi^*$ on whether a user likes to seek novelty or content with familiarity.}
\end{figure}
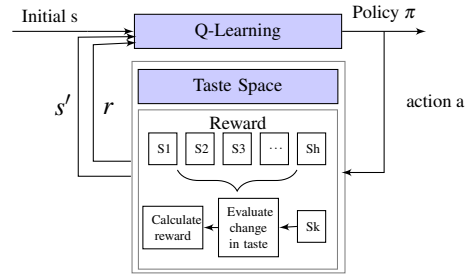
We use a model free reinforcement learning method called Q-learning \cite{kaelbling1996reinforcement} in which we do not need to know any transition probability model beforehand. With Q-learning, for each of the users, we learn state-action mapping that gives the expected utility of taking an action in a given state and in turn, it gives us an optimal policy. Cohen et.al. \cite{Cohen933} show that the exploratory preferences depend on user's state. We learn behaviors from users' taste states. In an episode, an action is taken, a reward is realized and an update to Q values is made. The magnitude of update depends on the learning rate and discount factor. We consider the state, s, at time t as user's tastes representation at time t. As illustrated in Figure 9, our policy learning agent learns an optimal policy in a taste state, $s$, with actions $a$, novelty seeking or content with familiarity, taking the user to a new state, $s'$ and it receives a reward. The cumulative reward is maximized, thereby, we learn an optimal policy which is a mapping of preferences with state as illustrated in the update equation below:  
\begin{equation}
\label{Q Learning}
Q(s,a) := Q(s,a) + \alpha[r + \gamma \max_{a'} Q(s',a') - Q(s,a)]
\end{equation}
where $\alpha$ is learning rate and $\gamma$ discount factor. 
Learning rate determines to what extent the values will be updated and discount factor determines the importance of future rewards. We set $\alpha$ as a decaying function of iterations, $\gamma$ as 0.9 and reward values are empirically set for users. Once the Q-learning converges, we get an optimal policy in a state, $\pi^*(s)$, given by 
\begin{equation}
\label{policy}
\pi^*(s) =  \arg\max_{a} Q^*(s,a)
\end{equation}
and this learned policy reflects users' novelty seeking behaviors.
\emph{Finally}, our presented model learns a crucial but hard to learn implicit user preferences like boredom or novelty seeking. Due to its flexible and modular design, the presented novelty learning method could play a significant role in making systems perceptive and trust worthy in a longer term.
\subsection{Evaluation Metrics}
Our measurement metrics are F1Score, Accuracy and Value of Personalization, as defined below:
\subsubsection{F1Score and Accuracy}
Given that $tp$ is true positive, $tn$ true negative, $fp$ false positive and $fn$ false negative, we can define F1Score and Accuracy as
\begin{align}
\label{F1Score and Accuracy}
P = tp/(tp + fp); R = tp/( tp + fn) \\
F1Score = (2 P R)/(P + R) \\
Accuracy = (tp + tn)/(tp + tn + fp + fn)
\end{align}
\subsubsection{Value of Personalization} 
User behaviors are individualistic and intrinsic which makes personalization of great value. For every user, the proposed model learns an optimal policy for novelty seeking preferences and we propose a metrics to understand the value of such personalization. Our intuition is to know how much a user would benefit going by a self policy than other user's policies. For every user, we capture F1Score using policy of other users other than self. We define this metrics $\Delta$ as 
\begin{equation}
\label{ValueOfPersonalization}
\Delta_{i} = Average(F1Score_{ij} - F1Score_{ii}) \forall i \neq j
\end{equation}
The greater this difference, the better the value of personalization for that particular user. 

\section{Experiment}
Our experiment is based on a publicly available users' music listening history dataset provided by last.fm \cite{Celma:Springer2010}. Last.fm is a popular music listening service helping users explore a variety of music, learn new music, connect with community and revisit previously listened tracks. Since this dataset has a temporal data, it forms a natural choice for our experiment requiring consumption diversity. Dataset \cite{lastfm} consists of complete music listening history of around 1K users from 2005 till 2009 with user id, time stamp, artist id, artist name, track id and track name for every user. The dataset statistics is provided in table 2.
\begin{table}[h!]
\centering
\begin{tabular}{|p{1.5cm}|p{1.5cm}|p{1cm}|p{1cm}|p{1.5cm}|} \hline
\textbf{Attribute} & Records & Unique Users & Artists & Unique Tracks\\ \hline
\textbf{Count} & 19,150,868 & 992 & 176,948 & 961,416 \\ \hline
\end{tabular}
\newline
\caption{Dataset Statistics}
\end{table}
\begin{algorithm}[h!]
\caption{Proximity Filtering and Novelty Learning}\label{TasteNoveltyLearn}
\begin{algorithmic}[]
\State Parameters: $K$ number of discovered tastes, w temporal window size, r reward values
\State x $\gets$ Input bag-of-words format \{pre-processing\}
\State $\alpha, \beta$ =  \textit{LDA}(x) \{get User Tastes and Taste Tracks dist.\}
\State $X_{m,t,s}$ = Input content consumption in temporal windows \{pre-processing\}
\State $X_{m,t}$ = Associate Taste Membership
\For {$j = 1:\textit{M}$} \{for each user\}
\\ \textit{Training}
\State Initialize RL Parameters
\While {not converged}
\State select action a with $\epsilon$-greedy policy from Q
\State take action a
\State s', r $\gets$ new state, reward
\State make Q-update \{Novelty Policy Q-Learning\}
\State s $\gets$ s'
\EndWhile
\State return policy
\\ \textit{Evaluate}
Evaluate on held out set
\EndFor
\end{algorithmic}
\end{algorithm}
Algorithm 1 presents the key steps involved.  In pre-processing and sessions management phases, we convert users' music listening data from a naive format to time series as well as bag-of-words format as required to discover tastes pool. An off-the-shelf package \cite{ldapython} is used for topic modeling. To assess a reasonable set of model parameters with stability, we computed the perplexity of the held out user set. The held-out set was unseen by the model during training. For a trained model, the perplexity of the unseen set is given by:
\begin{equation}
Perplexity(U_{test}) = exp\{-  \Sigma_{i=1}^{m} log (p(u_{i}| model ) ) /  \Sigma_{i=1}^{m} N_{i} \}
\end{equation}
where $log (p(u_{i}|model))$ is the log likelihood, calculated by fitting the trained model to the held-out set.
We varied number of topics from 1 to 40 and number of iterations for convergence as 10,100,1000 and 10000. We observed that the perplexity becomes stable beyond the number of iterations for convergence set as 100 iterations. We chose the number of topics to 20 which is at the beginning of a flat perplexity region and iterations as 1K because there is no gain in perplexity beyond these values. Moreover, we set up the temporal session window size to one hour empirically.  We varied it from 30 minutes - 1 week respectively. A larger window size does not reflect users' `in-moment' taste because users listen to tracks in proximity depending upon their current mood or state of mind. Using these parameters, we discover tastes to get tastes distribution per user and tracks distribution per taste and associate each of the temporal sessions with a taste. Therefore, users' items consumption history is transformed into a taste space.

The taste space representation is at an abstract level and we use it to learn novelty seeking Q-learning policy. Reinforcement learning requires learning rate and discount rate which are set as $1/(t^{0.65})$ and 0.9 respectively following several insights on convergence \cite{even2004learning}. For experimental simulation, the reward values are set in a range from -1 to 1 
as well as a small action cost. The Q-learning iterates until convergence and learns an optimal policy, whether a user is going to be novelty seeking or not given the current taste state of that user. We validate the learned policy using 80-20 training-validation split. Since the users' music listening data is in a time series format, this 80-20 split up is suitable when compared to other cross validation methods. We evaluate F1Score, Accuracy and Value of Personalization for every user.

In a supporting experiment to demonstrate users' quitting behavior and how taste representation can be beneficial, we use taste distribution per user from tastes discovery and create users' taste profiles. This is similar to the previous experiment except that instead of associating a taste, we associate a taste profile. Our aim, now, is to gain some insights into users' listening patterns with their taste profile similarities with previous sessions. Using this similarity score, we demonstrate that learning novelty in latent taste space is insightful in understanding users' content consumption behavior and knowing the potential customers who can quit.

To the best of our knowledge, there is no other computational method to learn inherent user tastes and dynamic behaviors like boredom or novelty seeking which are not directly observable. Therefore, a direct comparison with any method would be incomprehensible. However, to check our model, we compare our policy learning measures with that of Support Vector Machine with rbf kernel. Finally, we evaluate the value of personalization quantitatively.
\section{Results and Discussion}
We evaluated our model on the last.fm music listening dataset. We discuss user churn followed by novelty learning.
\subsection{User Quit and Novelty Seeking Behavior in Taste Space}
User churn is a major concern in a number of domains. Proactive knowledge of users who seek novelty is an important step to keep their trust in the recommender systems. We look at users' quitting behavior with respect to their taste profile sequences. We compute an average taste profile similarity score by comparing users' taste profile in the current session with that in the previous session. This measure is used to demonstrate our taste model's ability to learn user behavior using two example users - one who quit and stopped using the services within 449 days and other who continued to consume music listening services. Figure 10a demonstrates that the average number of tracks listened in a month by the `quitting' user continued to decline and the system was unable to act on this behavior because it had no clue if the user is seeking novelty or not. As shown in Figure 10b, the average profile similarity for this `quitting' user is lower with $\mu = 0.51$ and $\sigma = 0.15$  compared to the `continuing' user with $\mu = 0.89$ and $\sigma = 0.11$. It demonstrates that our novelty learning method in \emph{taste space} could play a pivotal role in understanding user behaviors proactively.
\begin{figure}[h!]
\centering
\begin{tabular}{c}
\includegraphics[width=1.0\linewidth,height=5.5cm]{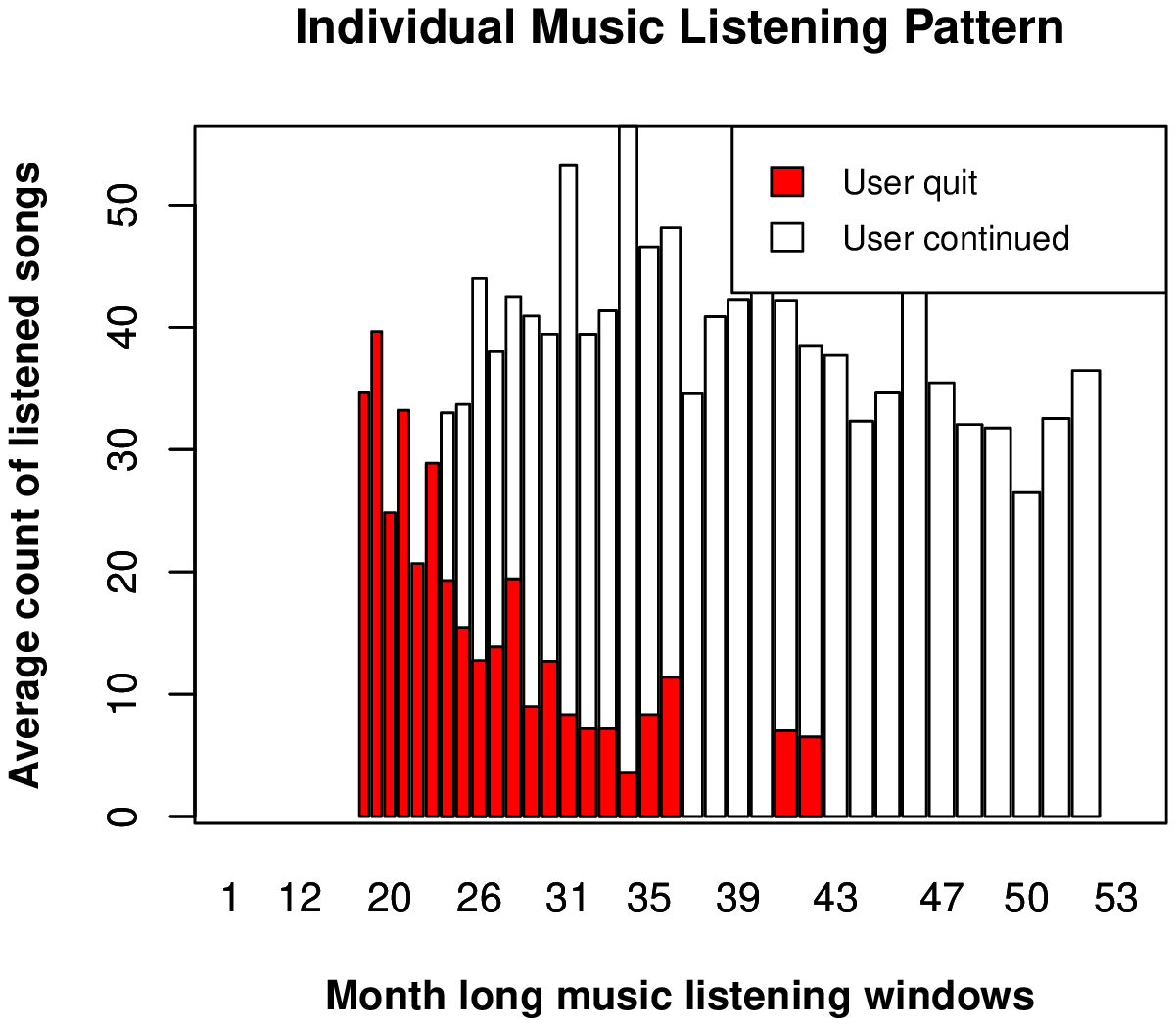}  \\
a. Songs listened in monthly windows \\
\includegraphics[width=1.0\linewidth,height=5.5cm]{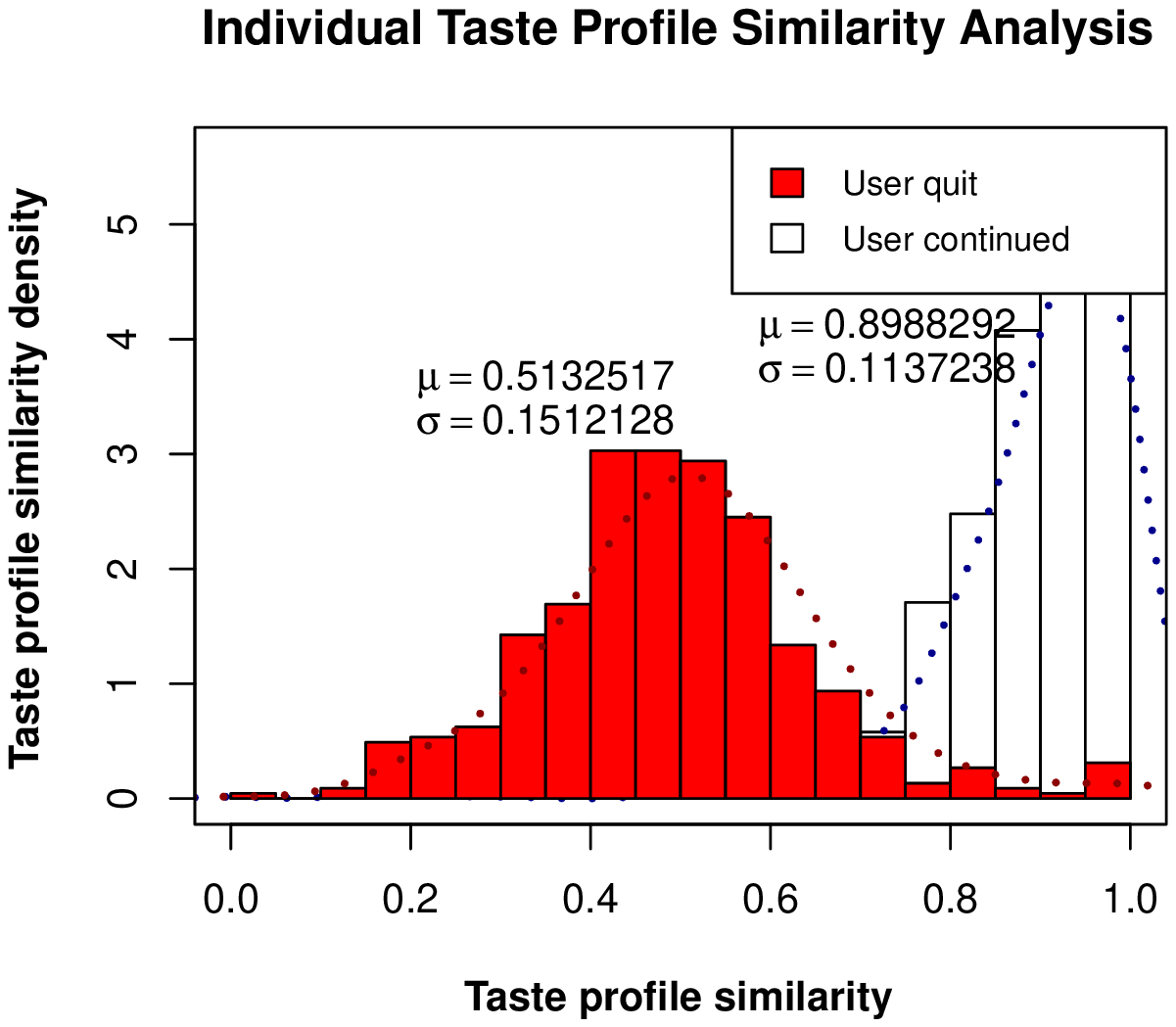}  \\
b. Taste profile similarity varies with the type of user \\
\includegraphics[width=1.0\linewidth,height=5.5cm]{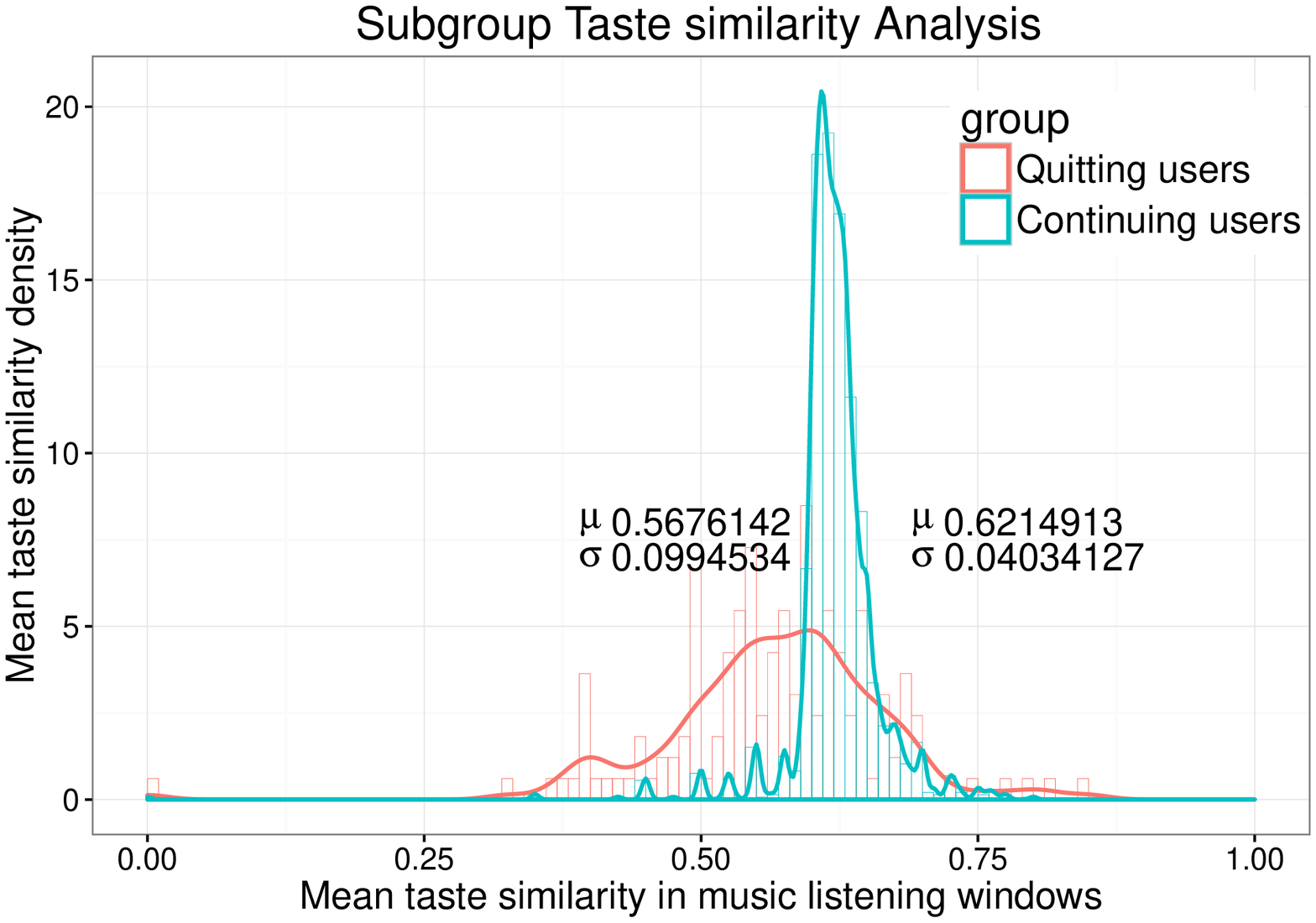}  \\
c. Two subgroups have different taste similarities
\end{tabular}
\caption{Taste space representation is adequate to proactively know which users are seeking novelty. \emph{a} demonstrates that the quitting user's listening activity continued to decline in comparison to the continuing user. \emph{b} compares taste profile similarity scores for the two users and demonstrates that quitting user was seeking more novelty. \emph{c} is subgroup level taste profile similarities comparison of quitting and continuing users. It demonstrates our ability to learn novelty in taste space.}
\label {fig:listening patterns}
\end{figure}

We also sought to understand an approximate user quiting behavior at a subgroup level. To achieve that, users were categorized based on their activities time period. We considered quitting users as the users whose activity time period from their first recorded activity to the last recorded activity was less than a year and their last recorded activity was at least one year prior to the end of the dataset time period. This ensured that we do not incorrectly categorize the users who started using the service very late. Similar to the individual user behavior analysis, we computed the taste similarity scores of users but at an aggregate level and demonstrate the subgroup level measures in Figure 10c. As evident, the quitting users has a lower mean taste profile similarity score implying a more exploratory behavior for this subgroup. It further establishes that the taste space representation is adequate and abstract enough to understand user's novelty seeking behavior.

\subsection{Novelty Policy Q-Learning}
Taste space representation of consumed content is used to learn users' novelty seeking behavior. Indirectly, learning novelty seeking behavior in taste space also means learning users' boredom with currently delivered content. Having such a representation and learning method is useful in content delivery systems involving human behaviors. With this view, we train our novelty learning agent, NoveltyLearn. We keep 20$\%$ of temporal data for validation and evaluate two measurement metrics F1Score and Accuracy.
\begin{table}[h!]
\centering
\renewcommand{\arraystretch}{1.1}
\begin{tabular}{|c|c|c|} \hline
$\downarrow$\textbf{Model} \textbf{Metric}$\rightarrow$ & \textbf{F1Score} &\textbf{Accuracy} \\ \hline
NoveltyLearn & 0.80326&0.87763 \\ \hline
SVM & 0.77700 & 0.83429 \\ \hline
\end{tabular}
\newline
\caption{Novelty Learning agent learns an optimal policy whether a user is seeking novelty or content with familiar items.}
\end{table}
The average evaluation F1Score of novelty seeking policy learner is 0.80326. Due to our model's ability to learn implicit tastes, any comparison with existing methods is incomprehensible but we compare average F1Score and Accuracy using single taste state representation of our method with a popular machine learning technique Support Vector Machines with RBF kernel using a longer taste history as input features. As shown in table 3, our learning agent, despite using only a single taste state with rewards, does better than SVM on an average by 2.6 points on F1Score and 4.3 points on Accuracy measure. The policy learner converges to an optimal policy, see Figure 11. 
\begin{figure}[h!]
\centering
\includegraphics[width=1.0\linewidth,height=4.5cm]{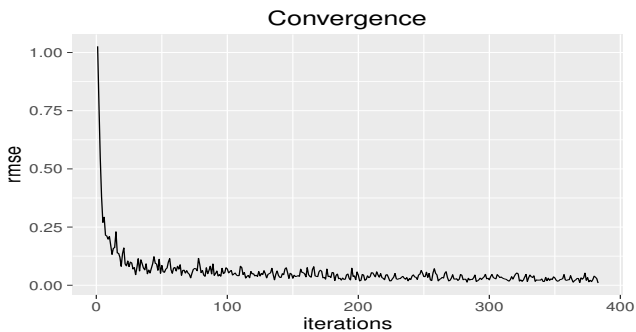}
\caption{Novelty learning agent converges to an optimal policy.}
\end{figure}

Evidently, the presented model is capable of learning user behavior in latent tastes space. The learned policy is useful to proactively understand user behavior and take action before they loose trust in system and quit. The novelty policy learning agent makes no consideration of user profile attributes like age, gender, location etc. which is a \emph{significant step} in establishing user profiles purely based on their consumption behavior. 

\subsection{Value of Personalization}
The value of personalization measure demonstrates how beneficial it is to consider user personalization. To evaluate it, we consider policies learned for users and apply them across other users except self on a 80-20 validation set. It means that for a user $i$, we evaluate the learner's policies according to all other users except $i$ to know how does it perform when it is not using its own policy. From this matrix form, the value of personalization is
\begin{figure}[h!]
\centering
\includegraphics[width=1.0\linewidth,height=4.25cm]{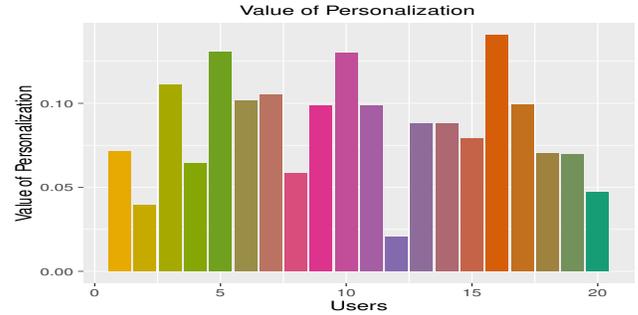}
\caption{Value Of Personalization demonstrates that personalization is beneficial.}
\end{figure}
 derived using the definition in equation 10. As shown in Figure 11, an overall average gain for a user using own policy is around 8 points. We observe that one user has a lower gain and it is attributed to a globally set parameter, length of sessions, to maintain evaluation consistency across users. Therefore, we state that it is beneficial to pursue personalization.
\subsection{Market Benefit}
Current recommender systems consider users' behavioral preferences as fixed and it is hard for computational systems to track unobservable behaviors like boredom. With an ability to learn such implicit behaviors, our method has a huge potential in a number of domains like music, movies etc. that rely heavily on user tastes and diversity of choices.
\section{Conclusion}
Our method fills a fundamental gap in learning implicit user behaviors. After an initial period of interaction with the system, users get bored with stale and repetitive content that does not match their implicit tastes and lose trust in the system. Therefore, an ability to learn users' tedium state is significant for user retention and satisfaction.

We presented a novel method called {\em collaborative proximity filtering} by leveraging patterns of consumption co-occurrence. The richness in understanding users' implicit and latent states that is provided by the taste model is not feasible in items space. We presented a learning agent that learns {\em novelty} seeking behavior in taste space. A proactive knowledge of such a user behavior is vital so that systems can take actions before a user actually churns and leaves the system. Our method provides a new way to profile users based on their taste learnt from the consumption proximity rather than the static attributes. Further, we demonstrated that personalization is useful and should not be ignored in recommender systems. The presented model brings complex psychological constructs within the purview of computation and a novel paradigm in creating a synergy between implicit behaviors and computational methods by learning from the data. 
\balance{}

\bibliographystyle{SIGCHI-Reference-Format}
\bibliography{novelty}

\end{document}